\documentclass[prd,twocolumn,letterpaper,showpacs,nofootinbib]{revtex4-2}

\usepackage{physics}
\usepackage{amsmath}
\usepackage{amssymb}  
\usepackage{graphicx}  
\usepackage{verbatim} 
\usepackage{color}    
\usepackage{subfigure}  
\usepackage{hyperref}  
\usepackage{mathrsfs}
\usepackage{bbold} 
\usepackage{cancel} 
\usepackage{empheq} 

\begin{document}

\title{How Can a Quantum Particle Be Found in a Classically Forbidden Region?}
\thanks{Journal Reference: The Physics Teacher, Vol. 63, January 2025, pp. 16-19. \url{https://doi.org/10.1119/5.0166989}}
\date{\today}
\author{Dennis E. Krause}
\author{Nikolai Jones }
\affiliation{Physics Department, Wabash College, Crawfordsville, IN 47933, USA}

\maketitle

Among the many perplexing results of quantum mechanics is one that contradicts a result from introductory physics: the possibility of finding a quantum particle in a region that would be forbidden classically by energy conservation~\cite{Knight}. An especially interesting example of this phenomenon with practical applications is quantum tunneling \cite{Knight 1168,Lincoln}. Here we investigate the reasons for this puzzling result by focusing on the difference between how quantities like kinetic and potential energy are represented mathematically in classical and quantum mechanics. In quantum mechanics, physical observables, like energy, are represented by operators rather than real numbers. The consequences of this difference will be illustrated explicitly using a toy model in which the kinetic and potential energy operators are represented by $2 \times 2$ matrices, which do not commute like their classical analogs. This model will then illustrate how classically perplexing results, like a quantum particle being found in the forbidden region, can arise.

\
When first encountering potential energy in introductory physics, it is often useful to create an energy graph \cite{Knight 244}. For a particle moving in one dimension along the $x$-axis, one typically plots the potential energy $V = V(x)$ as a function of $x$ and total energy, 
\begin{equation}
K + V = E,
\end{equation}
which is a constant for an isolated system. The particle's kinetic energy $K$ when the particle is at position $x$ is the difference: $K = E - V(x)$. Figure 1 depicts an energy graph for a particle confined by a 1-dimensional simple harmonic oscillator (SHO) potential energy,
\begin{equation}
V(x)=\frac{1}{2}kx^{2}.	
\end{equation}

\begin{figure}[h]
\includegraphics[width=3.2in]{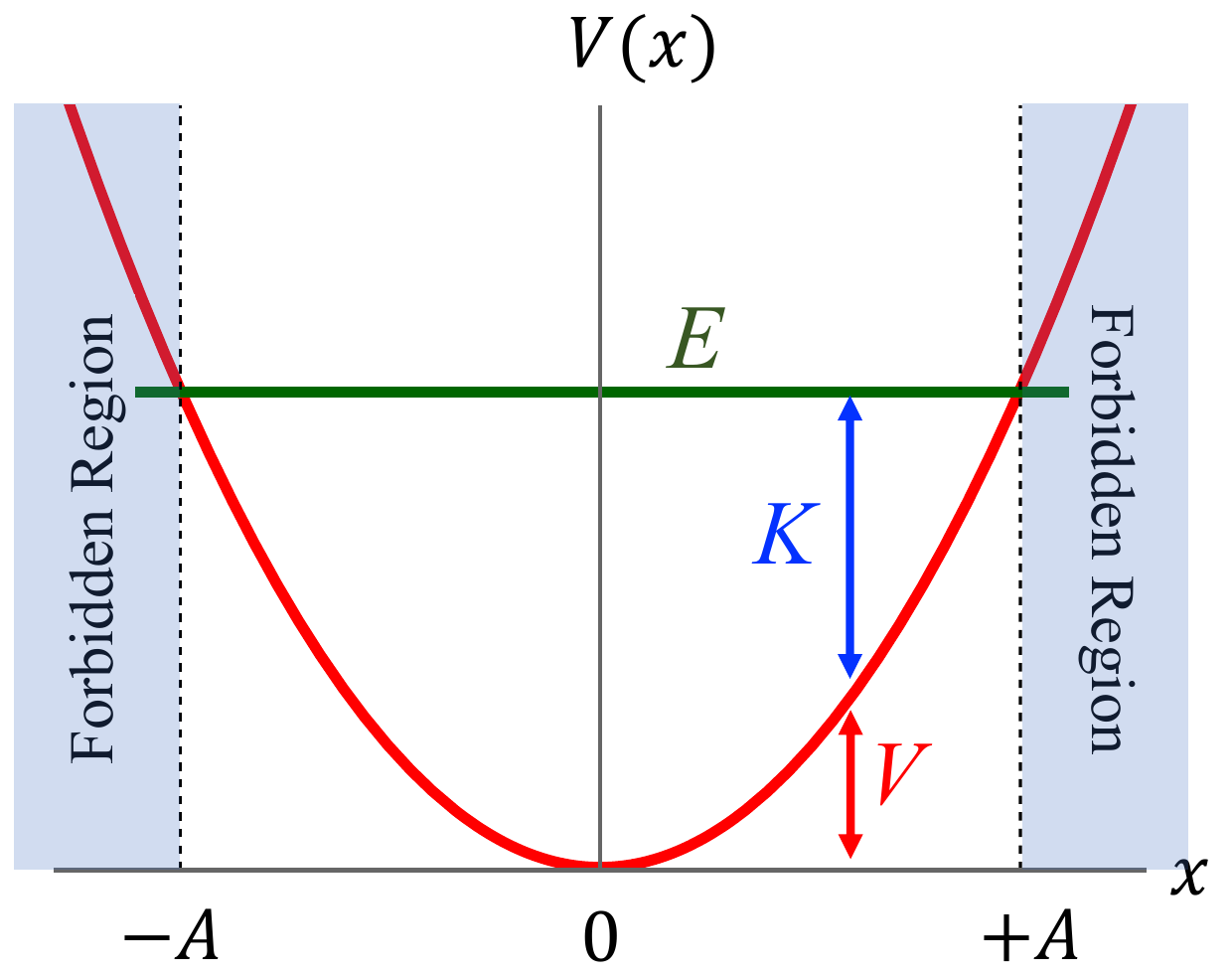}
\caption{An energy graph for a particle confined to a one-dimensional simple harmonic oscillator potential $V(x)$, Eq.~(2), with total energy $E = K + V$, where $K$ is the particle's kinetic energy. The particle oscillates in the region $|x| \leq A$, where $A$ is the amplitude of oscillation; it is excluded from the shaded ``Forbidden Region'' by energy conservation.}
\end{figure}

Since the simple harmonic oscillator particle's kinetic and potential energies are both non-negative, the particle is confined by energy conservation to move only in the region $|x|\leq A$, where the amplitude $A=\sqrt{2E/k}$.  Assuming the potential energy given by Eq.~(2), the particle would need to have negative kinetic energy to be in the ``forbidden region,'' $|x|>A$, which is impossible.  However, when we study the same problem of a particle confined to a one-dimension SHO potential using quantum mechanics, we find something remarkable---there is a significant probability of finding the particle in the classically forbidden region!  

To better understand how the forbidden region does not apply in quantum mechanics, let's focus on the ground state, the lowest energy state that a quantum particle confined to a SHO potential can have \cite{Knight 1162}. In classical mechanics, the lowest energy is $E = 0$, when the particle is at rest at the equilibrium position $x = 0$ so that $K = V= 0$. However, this state is forbidden by the Heisenberg uncertainty principle, which says that a quantum particle cannot simultaneously have a well-defined position and momentum \cite{Knight 1131,Hobson}. Classically, the ground state has both: $x=0$ and $p_x=0$.  Instead, a quantum particle in its ground state has a non-zero energy $E_{\rm ground}=\hbar\omega/2$ .  Therefore, if we are to compare the classical and quantum SHOs, let's assume that their particles both have the same total energy $E =E_{\rm ground}$.  Now if we calculate the probability density of finding the particle at a position $x$ for each case, we obtain the results shown in Fig.~2.

\begin{figure}[h]
\includegraphics[width=3.2in]{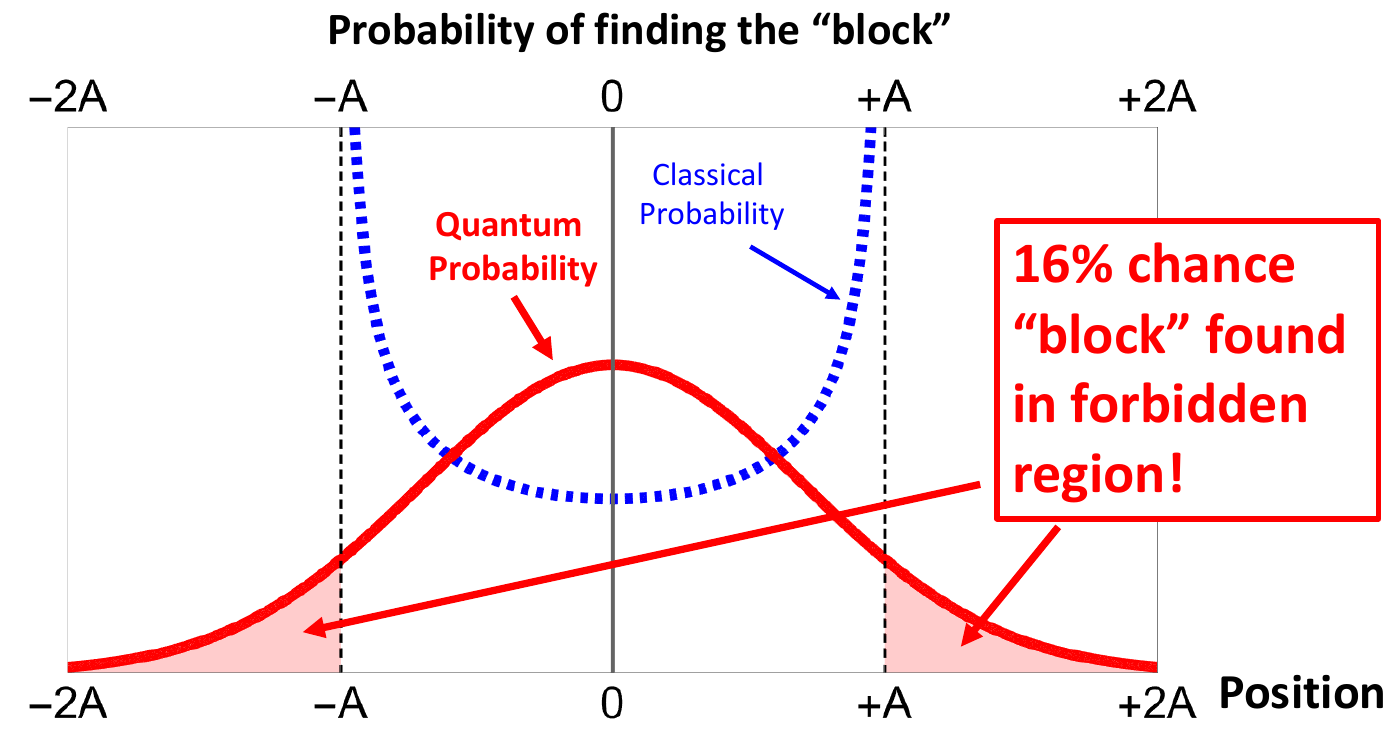}
\caption{Graphs of the classical and quantum probability densities of finding a particle at position $x$ if both classical and quantum particles have the same total energy $E = E_{\rm ground} = \hbar\omega/2$. The shaded area under the curve represents the probability of finding the particle in the forbidden region.}
\end{figure}

The graphs in Fig. 2 reveal the stark contrast between the classical and quantum cases.  There we find the classical particle is confined in the region $|x|\leq A$ as expected, and that the particle is most likely to be found near $|x|=A$.  Why?  Classically, the 
particle is most likely to be found where it is traveling slowest (has the smallest kinetic energy), which is near $|x|=A$; it is least likely to be found where it has the greatest kinetic energy, near the equilibrium position $x=0$.  By contrast, the graph of the quantum position probability density in Fig. 2 reveals the quantum particle in its ground state displays the opposite behavior.  It is most likely to be found near the equilibrium position, and, more interestingly, there is a significant (16\%) probability of finding the particle in the classically forbidden region $|x|>A$.  Since the potential energy is assumed to be positive, does this mean a quantum particle can have negative kinetic energy in the forbidden region?  It turns out quantum mechanics may be unintuitive, but it is not {\em that} unintuitive---kinetic energy in quantum mechanics is non-negative just as in classical mechanics.  

A way to understand how a quantum particle can be found in the forbidden region when both kinetic and potential energies are positive can be found in an analogous kinematics problem from introductory physics which is probing a student's understanding of distance and displacement:
\begin{quotation}
\em
On a sunny spring day, Alicia decides to take a hike. First, she walks a distance $d_1=100$ meters east.  Then she changes direction and walks a similar distance $d_2=100$ meters.  She now finds herself only 50 meters from her starting point. 
\end{quotation}
How is this possible? After all,
\begin{equation}
d_1+d_2=100+100\neq50.		
\end{equation}
Does this imply that one of the distances is somehow negative?  Of course not. The answer is that distances, which are always positive, do not add like real numbers to obtain the net displacement magnitude (Fig. 3).  Rather, the parts of Alicia's journey are represented mathematically by vector displacements, $\Delta \vec{r}_{1}$ and  $\Delta \vec{r}_{2}$ which add according to vector addition:
\begin{equation}
 \Delta \vec{r}_{1}+ \Delta \vec{r}_{1}= \Delta \vec{r}_{\rm tot}.
\end{equation}
\begin{figure}[h]
\includegraphics[width=3.2in]{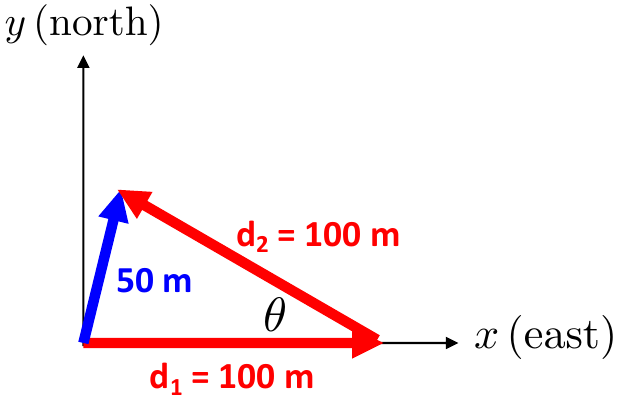}
\caption{The vectors representing the displacements of the two 100-m-long segments of Alicia's hike are drawn.  The distance from the origin after the final segment (50 m) is the magnitude of the vector sum (total displacement), not the sum of the segment distances.}
\end{figure}
The final distance from her starting point is determined from the magnitude of the total displacement $|\Delta \vec{r}_{\rm tot}|$, which, using the law of cosines, is given by
\begin{equation}
\Delta \vec{r}_{\rm tot}| = \sqrt{(d_{1})^{2} + (d_{2})^{2} - 2d_{1}d_{2}\cos\theta},
\end{equation}
where $\theta$ is the angle shown in Fig. 3.  We see that only when $\theta=180^{\circ}$ does the total displacement magnitude equal the sum of the segment distances.

The intended lesson from the Alicia hike problem is that using the wrong mathematical representation, representing path displacements by real numbers (distances) instead of vectors, leads to perplexing and incorrect results.  This problem may seem far removed from trying to understand how a quantum particle can be in the forbidden region, but we will see that an analogous flaw in reasoning is making the quantum problem seem more mysterious than it actually is.

We start by returning to the energy equation, Eq.~(1), which assumes that the total energy is the sum of two real numbers, the kinetic and potential energies.  However, in 1925, Werner Heisenberg recognized that a new mathematical representation of energy was needed to explain atomic spectra and other phenomena that had been perplexing physicists at that time \cite{Matrix}.  Instead of representing the kinetic, potential, and total energies by real numbers, it was shown that these quantities need to be represented by linear operators, $\hat{K}$, $\hat{V}$, and $\hat{E}$ respectively. (Here the hats over symbols mean the quantities are operators, not unit vectors.)  Eq.~(1) then becomes
\begin{equation}
 \hat{K} + \hat{V} = \hat{E}.
\end{equation}
The kinetic energy operator depends on the momentum operator $\hat{P}_{x}$ as
\begin{equation}
\hat{K} = \frac{\hat{P}_{x}^2}{2m},
\end{equation}
while the potential energy operator depends on the position operator $\hat{X}$ as
\begin{equation}
\hat{V} = V_{\rm SHO}(\hat{X}) = \frac{1}{2}k\hat{X}^2.
\end{equation}

In wave mechanics, the 1-dimensional position and momentum operators are given by \cite{Momentum}
\begin{equation}
\hat{X} = x, \phantom{space}   \hat{P}_{x} = -i\hbar\frac{\partial}{\partial x}, 
\end{equation}
where $\hbar = h/(2\pi)$ is the rationalized Planck's constant.  The crucial point is that the position and momentum operators don't commute---the order of their multiplication matters:
\begin{equation}
\hat{X}\hat{P}_{x} - \hat{P}_{x}\hat{X} = i\hbar.
\end{equation}
We can see this using Eq.~(9) and the chain rule for differentiation: 
\begin{equation}
\hat{X}\hat{P}_{x} = x\left(-i\hbar\frac{\partial}{\partial x}\right),
\end{equation}
\begin{equation}
\hat{P}_{x}\hat{X} = -i\hbar\frac{\partial}{\partial x}\,x = -i\hbar +x\left(-i\hbar\frac{\partial}{\partial x}\right).
\end{equation}
A consequence of this is the Heisenberg uncertainty principle \cite{Uncertainty},  which prevents the ground state of the SHO from having vanishing energy.

To illustrate how the new mathematical operator representation changes how one deals with the energy equation (6) without the technicalities of calculus, let's consider a toy model where the kinetic and potential energy operators are instead represented by $2\times 2$ matrices.  (After all, Heisenberg's formulation of quantum mechanics is often called ``matrix mechanics.'')  Let the kinetic energy operator be given by the matrix
\begin{equation}
\hat{K} = \begin{pmatrix}
0 & 0 \\
0 & 2
\end{pmatrix}.
\end{equation}

In quantum mechanics, when an operator matrix is written in diagonal form like this, the diagonal matrix elements are the possible outcomes of a measurement result \cite{Mathematical}.   These elements are the {\em eigenvalues} of the matrix.  In this case, a measurement of the kinetic energy of the particle would have only two possible outcomes, $K = 0$ or $K = 2$.  For the potential energy operator in our toy model, let us write it as
\begin{equation}
 \hat{V} = \begin{pmatrix}
1 & 1 \\
1 & 1
\end{pmatrix}.
\end{equation}

This matrix is not written in diagonal form.  However, using the standard methods of linear algebra, we would find that its eigenvalues are also 0 or 2 so a measurement of potential energy would only yield either $V=0 $ or $V=2$.  [We'll see why we don't write Eq.~(14) like Eq.~(13) in a moment.]

Like the actual kinetic and potential operators for the simple harmonic oscillator, the toy model operators don't commute $\hat{K}\hat{V} \neq \hat{V}\hat{K}$:
\begin{equation}
\hat{K}\hat{V} = 
\begin{pmatrix}
2 & 0 \\
0 & 0
\end{pmatrix}
\begin{pmatrix}
1 & 1 \\
1 & 1
\end{pmatrix}
=
\begin{pmatrix}
2 & 2 \\
0 & 0
\end{pmatrix}
\end{equation}
\begin{equation}
\hat{V}\hat{K} = 
\begin{pmatrix}
1 & 1 \\
1 & 1
\end{pmatrix}
\begin{pmatrix}
2 & 0 \\
0 & 0
\end{pmatrix}
=
\begin{pmatrix}
2 & 0 \\
2 & 0
\end{pmatrix}.
\end{equation}
This is a general property of matrices---the order of multiplication matters.  

What is the consequence of the non-commutativity of the operators?  A mathematical theorem from linear algebra tells us that if two matrices do not commute, they cannot be simultaneously diagonalized \cite{Griffiths}.   This means that in our toy model, the system cannot simultaneously have definite values of kinetic and potential energy as in the real SHO system.  In the case of the position and momentum operators not commuting, Eq.~(10), we find that the system cannot simultaneously have definite values of position and momentum, which is the meaning of the Heisenberg uncertainty principle.

Let's now look at the total energy operator given by Eq.~(6).  Inserting the toy kinetic and potential energies, we find the matrix representing the total energy of the toy system is \cite{equation}
\begin{equation}
\hat{E} = \hat{K} + \hat{V} = 
\begin{pmatrix}
2 & 0 \\
0 & 0
\end{pmatrix}
+
\begin{pmatrix}
1 & 1 \\
1 & 1
\end{pmatrix}
=
\begin{pmatrix}
3 & 1 \\
1 & 1
\end{pmatrix}.
\end{equation}
The eigenvalues of the matrix $\hat{E}$ are 0.59 and 3.41, so if we measure the model system's total energy, the only possible values that can be found are $E=0.59$ and $E=3.41$.  Notice that these two values do not correspond to the sum of the eigenvalues of the toy model kinetic and potential energy eigenvalues, which would be $0 + 0 = 0$ or $2 + 2 = 4$ \cite{Operator}.   We now have everything we need to address the issue at hand.

Suppose we measure the total energy of our toy system and obtain $E = 0.59$.  Immediately afterward, we measure the system's kinetic energy.  What is the probability of obtaining $K = 2$?  Classically, this is impossible since the kinetic energy cannot be greater than the total energy if both kinetic and potential energies are non-negative.  However, according to the rules of quantum mechanics, we find for this system that the probability of obtaining $K = 2$ is 15\%.  (The probability of getting the other possibility $K = 0$ is 85\%.)  We find that a part ($K$) is larger than the total ($E$)!

When we use Eq.~(1) in classical mechanics, we are assuming that the kinetic and potential energies are represented by real numbers and they simultaneously have definite values, so we can add them.  However, when we attempt to do this in quantum mechanics, we are making the same mistake an introductory student makes in the Alicia hike problem attempting to add the distances of each segment to obtain the distance from her starting point.  Rather, the student should be representing the path segments by vectors.  Similarly, in quantum mechanics, experiments tell us that energies are represented by linear operators, not real numbers, which don't usually commute.  Eq.~(1) needs to be replaced by the operator equation, Eq.~(6).  The results of experimental measurements are now obtained from the eigenvalues of these operators. The mathematics of linear operators, linear algebra, combined with the rules of quantum mechanics can often give us puzzling results when viewed from the perspective of classical mechanics which uses a different mathematical representation for energy (real numbers).  

Like students encountering vectors for the first time, requiring them to learn the machinery of coordinate systems, components, etc., a student discovering quantum mechanics is faced with new mathematics and rules for calculation which take time to get used to but lead to results that agree with experiments \cite{Susskind}.   Quantum particles can be in the forbidden region because the classical assumptions that make the region forbidden (kinetic and potential energies are represented by real numbers that have simultaneously definite values) simply don't apply.  This is also why a quantum particle does not have simultaneously definite values of position and momentum---the Heisenberg uncertainty principle! Only when operators representing physical observables commute will they behave more like their classical counterparts.


\begin{thebibliography}{99}
\bibitem{equation} This equation helps explain why we didn't write $\hat{V}$ in diagonal form, because we would have to use a different basis because of the non-commutativity theorem.  When adding matrices, we must use the same basis for both, in the same way that adding the components of two vectors requires using the same coordinate system for both.
\bibitem{Operator} One can show that if the kinetic and potential operators commute, then the total energy eigenvalues would be the sum of the kinetic and potential energy eigenvalues.  This corresponds to a classical-like system, where the total energy is the sum of the kinetic and potential energy values.  However, the Heisenberg uncertainty principle prevents this from happening.
\end{thebibliography}
\end{document}